\documentclass[12pt]{iopart}
\usepackage{graphicx}
\def\be{\begin{equation}}
\def\ee{\end{equation}}
\def\bea{\begin{eqnarray}}
\def\eea{\end{eqnarray}}

\begin{document}

\title{$C^3$ matching for asymptotically flat spacetimes}

\author{Antonio  C. Guti\'errez-Pi\~{n}eres}
\address{Escuela de F\'\i sica, Universidad Industrial de Santander, CP 680002,  Bucaramanga, Colombia\\
Instituto de Ciencias Nucleares, Universidad Nacional Aut\'onoma de M\'exico,
 AP 70543,  M\'exico, DF 04510, M\'exico\\   
 Facultad de Ciencias B\'asicas, Universidad Tecnol\'ogica de Bol\'ivar,  CP 131001,  Cartagena, Colombia}
\ead{acgutier@uis.edu.co} 

\author{Hernando Quevedo}
\address{Instituto de Ciencias Nucleares, Universidad Nacional Aut\'onoma de M\'exico, AP 70543, Ciudad de M\'exico 04510, Mexico\\
Dipartimento di Fisica and ICRA, Universit\`a di Roma "Sapienza", I-00185, Roma, Italy\\
 Institute of Experimental and Theoretical Physics, 
	Al-Farabi Kazakh National University, Almaty 050040, Kazakhstan}
\ead{quevedo@nucleares.unam.mx}


\begin{abstract}
We propose a criterion for finding the minimum distance at which an interior solution of Einstein's equations can be matched with an exterior asymptotically flat solution. It is based upon the analysis of the eigenvalues of the 
Riemann curvature tensor and their first derivatives, implying  $C^3$ differentiability conditions. 
The matching itself is performed by demanding continuity of the curvature eigenvalues across the matching surface. 
We apply the $C^3$ matching approach to spherically symmetric perfect fluid spacetimes and obtain the physically meaningful condition that density and pressure should vanish on the matching surface. Several perfect fluid solutions in Newton and Einstein gravity are tested. 
\end{abstract}

%
%
%
%
%

\section{Introduction}
\label{sec:int}

General relativity is a theory of the gravitational interaction and,  in particular, should describe the gravitational field of relativistic compact objects. In this case, the spacetime can be split 
into two different parts, namely, the interior region described by an exact solution $g_{\mu\nu}^-$ of Einstein's equations with a physically reasonable energy-momentum tensor and the exterior region, 
which corresponds to an exact vacuum solution $g_{\mu\nu}$. This implies that the spacetime $M$ can be considered as split into two regions $(M^-,g_{\mu\nu}^-)$ and $(M^+,g_{\mu\nu}^+)$ with a special 
hypersurface $\Sigma$ at which the two regions should be matched. In the case of compact objects, $\Sigma$ should be identified with the surface of the object, i.e., it is a time-like hypersurface.  
It then follows that at $\Sigma$ certain matching conditions should be imposed in order for the spacetime to be well defined. 

Two sets of matching conditions are commonly used in the literature. The Darmois conditions \cite{darmois1927equations} demand that the first and second fundamental forms (the intrinsic metric and the 
extrinsic curvature) be continuous across $\Sigma$.  The Lichnerowicz conditions state that the metric and all its first derivatives must be continuous across $\Sigma$ in ``admissible'' coordinates 
that traverse $\Sigma$. The Darmois conditions are expressed in terms of tensorial quantities and hence they can be considered as a covariant formulation of the matching problem. In the case of the 
Lichnerowicz conditions, the term ``admissible'' coordinates is used, indicating that the choice of a coordinate system is essential. The equivalence between Darmois and Lichnerowicz conditions can be 
proved by using Gaussian normal coordinates. This proof allows one also to precise the concept of ``admissible'' coordinates which are then defined as coordinates related to Gaussian normal 
coordinates by means of ($C^2_\Sigma, C^4$) transformations \cite{lake2017revisiting}. However, as pointed out by Israel in \cite{israel1966singular}, the explicit form of the matching conditions are 
of limited utility, since ``admissible'' coordinates usually are not the most convenient for handling the matching problem in practice. 

An alternative approach in which the extrinsic curvature is not necessarily continuous across $\Sigma$ was proposed by Israel. An effective energy-momentum tensor which determines a thin shell is 
defined in terms of the difference of the extrinsic curvature evaluated inside and outside the hypersurface $\Sigma$. This means that $\Sigma$ can now be interpreted as a thin shell that separates 
$M^+$ and $M^-$, is part of the entire spacetime $M$  and as such plays an important role in the determination of the spacetime  dynamics.

In all the $C^2$ matching conditions described above, it is important to know a priori the location of $\Sigma$. Although in the case of compact objects, we can identify $\Sigma$ with the surface of 
the body, in general, it is not easy to find the equation that determines the surface and, if possible, it often is not given in the ``admissible'' coordinates that are essential for treating the 
matching problem. This is probably the reason why the matching conditions have been applied so far only in cases characterized by very high symmetries.

In this work, we propose to use a $C^3$ criterion to find information about the location of the hypersurface $\Sigma$. It is defined in terms of the eigenvalues of the Riemann curvature tensor which 
are invariant quantities. The  idea is simple. Since the curvature tensor is a measure of the gravitational interaction, the curvature eigenvalues provides us with an invariant measure of the 
gravitational interaction. Since for a compact object, one expects the spacetime to be asymptotically flat, the curvature eigenvalues should vanish at spatial infinite, the behavior of the eigenvalues 
approaching the gravitational source could give some information about its borders. Here, we implement this simple idea in an invariant manner and show its applicability in the case of several exact 
solutions of Einstein's equations.

This paper is organized as follows. In Sec. \ref{sec:curv}, we use Cartan's formalism to investigate the general form a curvature tensor that satisfies Einstein's equations with a perfect fluid source. Moreover, we find the general form of the curvature eigenvalues and derive some identities relating them. In Sec. \ref{sec:rep}, we review the definition of repulsive gravity in terms of the curvature eigenvalues. This definition is then used in Sec. \ref{sec:mat} to propose the $C^3$ matching approach, whose objective is to perform the matching in such a way that the effects of repulsive gravity cannot be detected. We also apply the method to spherically symmetric solutions in Newton and Einstein gravity, obtaining physically meaningful results. Finally, in Sec. \ref{sec:con}, we discuss our results and propose some tasks for future investigation.


\section{Curvature eigenvalues and Einstein equations }
\label{sec:curv}

Our approach is based upon the analysis of the behavior of the curvature eigenvalues. There are different ways to determine these eigenvalues \cite{stephani2009exact}. Our strategy is to use local 
tetrads and differential forms. From the physical point of view, a local orthonormal tetrad is the simplest and most natural choice for an observer in order to perform local measurements of time, 
space, and gravity. Moreover, once a local orthonormal tetrad is chosen, all the quantities related to this frame are invariant with respect to coordinate transformations. The only freedom remaining 
in the choice of this local frame is a Lorentz transformation. So, let us choose the orthonormal tetrad as 
\be
ds^2 = g_{\mu\nu} dx^\mu\otimes dx^\nu= \eta_{ab}\vartheta^a\otimes\vartheta^b\ ,
\ee
with $\eta_{ab}={\rm diag}(-1,1,1,1)$, and $\vartheta^a = e^a_{\ \mu}dx^\mu$. The first
\be
d\vartheta^a = - \omega^a_{\ b }\wedge \vartheta^b\ ,
\ee
and second Cartan equations
\be
\Omega^a_{\ b} = d\omega^a_{\ b} + \omega^a_{ \ c} \wedge \omega^c_{\ b} = \frac{1}{2} R^a_{\ bcd} \vartheta^c\wedge\vartheta^d
\ee
allow us to compute the components of the Riemann curvature tensor in the local orthonormal frame.

It is possible to represent the curvature tensor as a  (6$\times$6)-matrix by introducing the bivector indices $A,B,...$ which encode 
the information of two different tetrad indices, i.e., $ab\rightarrow A$. We follow the convention proposed in \cite{misner2017gravitation} which establishes the following correspondence between 
tetrad and bivector indices
\be
01\rightarrow 1\ ,\quad 02\rightarrow 2\ ,\quad 03\rightarrow 3\ ,\quad 23\rightarrow 4\ ,\quad 31\rightarrow 5\ ,\quad 12\rightarrow 6\ .
\ee
Then, the Riemann tensor can be represented by the symmetric matrix ${\bf R}_{AB} = {\bf R}_{BA}$ with 21 components. The first Bianchi identity $R_{a[bcd]}=0$, which in bivector representation reads 
\be 
{\bf R}_{14}+{\bf R}_{25}+{\bf R}_{36}=0\ ,
\ee
reduces the number of independent components to 20.

Einstein's equations with cosmological constant
\be
R_{ab} - \frac{1}{2} R \eta_{ab} + \Lambda \eta_{ab} = \kappa T_{ab}\ ,\quad R_{ab} = R ^c _{\ acb}\ ,
\ee
can be written explicitly in terms of the components of the curvature tensor in the bivector representation, resulting in a set of ten algebraic equations that relate the components ${\bf R}_{AB}$.
This means that only ten components ${\bf R}_{AB}$ are algebraic independent which can be arranged in the $6\times 6$ curvature matrix 
\be \label{eq: CurvatureTensor}
{\bf R}_{AB}=\left(
         \begin{array}{cc}
           {\bf M}_1 & {\bf L} \\
           {\bf L} & {\bf M}_2 \\
         \end{array}
       \right),
\ee
where 
$$
{\bf L} =\left(
         \begin{array}{ccc}
			{\bf R}_{14}  & 	{\bf R}_{15}  & {\bf R}_{16} \\
			{\bf R}_{15} - \kappa T_{03} &  {\bf R}_{25} &  {\bf R}_{26}  \\
			{\bf R}_{16} + \kappa T_{02}  &   \quad {\bf R}_{26}  - \kappa  T_{01} & \quad - {\bf R}_{14} -	{\bf R}_{25}   \\
         \end{array}
       \right), 
      $$ 
and 
${\bf M}_1$ and  ${\bf M}_2$ are $3\times 3$ symmetric matrices
$$
{\bf M}_1=\left(
         \begin{array}{ccc}
			{\bf R}_{11} &  \quad	{\bf R}_{12} & {\bf R}_{13} \\
			{\bf R}_{12} & \quad {\bf R}_{22} &  {\bf R}_{23} \\
			{\bf R}_{13} & \quad   {\bf R}_{23} &  \quad - {\bf R}_{11}   -	{\bf R}_{22} - \Lambda  {+} \kappa \left(\frac{T}{2} +T_{00}\right)  \\
         \end{array}
       \right),
       $$

$$
{\bf M}_2 = 
\small{   \left( \\
         \begin{array}{ccc}
			-{\bf R}_{11} + \kappa \left(\frac{T}{2} +T_{00}-T_{11} \right)   &  {-} { \bf R}_{12} - \kappa T_{12}   & - {\bf R}_{13} - \kappa T_{13} \\
			{-} { \bf R}_{12} - \kappa T_{12}   &   -{\bf R}_{22} + \kappa \left(\frac{T}{2} +T_{00}-T_{22} \right)   &  - {\bf R}_{23} - \kappa T_{23} \\
		- {\bf R}_{13} - \kappa T_{13}    &    - {\bf R}_{23} - \kappa T_{23}   &   {\bf R}_{11} +	{\bf R}_{22} + \Lambda {-} \kappa T_{33}   \\
         \end{array}
       \right)   },
 $$     
with $T=\eta^{ab}T_{ab}$. 
This is the most general form {of} a curvature tensor that satisfies Einstein's equations with cosmological constant and arbitrary energy-momentum tensor. The traces of the matrices entering the final form of the curvature turn out to be of particular importance. First, 
the matrix ${\bf L}$ is traceless by virtue of the Bianchi identities, as shown above. Moreover, for the remaining matrices we obtain
\be
{\rm Tr}({\bf M_1}) = - \Lambda + \kappa \left(\frac{T}{2} +T_{00}\right) \ , \quad
{\rm Tr}({\bf M_2}) = + \Lambda + \kappa T_{00}\ ,
\ee
so that 
\be
{\rm Tr}({\bf R_{AB}}) = \kappa \left(\frac{T}{2} +2 T_{00}\right)\ .
\ee
We see that these traces depend on the components of the energy-momentum tensor only. Then, in the particular case of vacuum fields with vanishing cosmological constant, the curvature matrix reduces to 
\be
{\bf R}_{AB}=\left(
         \begin{array}{cc}
           {\bf M} & {\bf L} \\
           {\bf L} & -{\bf M} \\
         \end{array}
       \right), \quad
			{\bf M}=\left(
         \begin{array}{ccc}
			{\bf R}_{11} & 	{\bf R}_{12} & {\bf R}_{13} \\
			{\bf R}_{12} &  {\bf R}_{22} &  {\bf R}_{23} \\
			{\bf R}_{13} &  {\bf R}_{23} &   {\bf R}_{33}  \\
         \end{array}
       \right) \ ,
\ee
and the $3\times 3$ matrices $L$ and $M$ are symmetric and trace free,
\be
{\rm Tr}({\bf L})=0\ , \qquad {\rm Tr}({\bf M})=0\ .
\ee

For later use, we also consider the case of a perfect fluid energy-momentum tensor with density $\rho$ and pressure $p$ 
\be
T_{ab} = (\rho+p)u_a u_b + p \eta_{ab} \ ,
\ee
where $u_a$ is the four-velocity of the fluid which for simplicity can always be chosen as the comovil velocity $u^a=(-1,0,0,0)$. Then,
\be
T_{ab}= {\rm diag}(\rho,p,p,p)\ .
\ee

The curvature matrix for a perfect fluid  is then  given by   Eq.(\ref{eq: CurvatureTensor})  with
$$
{\bf L} =\left(
         \begin{array}{ccc}
			{\bf R}_{14} & 	{\bf R}_{15} & {\bf R}_{16} \\
			{\bf R}_{15}  &  {\bf R}_{25} &  {\bf R}_{26} \\
			{\bf R}_{16}   &  {\bf R}_{26}  & - {\bf R}_{14} -	{\bf R}_{25}   \\
         \end{array}
       \right), 
\label{pf1}
$$

$$
{\bf M}_1=\left(
         \begin{array}{ccc}
			{\bf R}_{11} & 	{\bf R}_{12} & {\bf R}_{13} \\
			{\bf R}_{12} &  {\bf R}_{22} &  {\bf R}_{23} \\
			{\bf R}_{13} &  {\bf R}_{23} & - {\bf R}_{11} -	{\bf R}_{22} - \Lambda {+} \frac{\kappa}{2} \left( 3 p + \rho \right)  \\
         \end{array}
       \right),
\label{pf2}
$$

$$
{\bf M}_2= \left(
         \begin{array}{ccc}
			-{\bf R}_{11} + \frac{\kappa}{2} (\rho +p)   & {-} { \bf R}_{12}  & - {\bf R}_{13}  \\
			{-} { \bf R}_{12}  &   -{\bf R}_{22} + \frac{\kappa}{2} (\rho +p)    &  - {\bf R}_{23}  \\
		- {\bf R}_{13}  & - {\bf R}_{23}   &  {\bf R}_{11} +	{\bf R}_{22} + \Lambda  - \kappa p   \\
         \end{array}
       \right) .
\label{pf3}
$$

The eigenvalues of the curvature tensor correspond to the eigenvalues of the matrix ${\bf R}_{AB}$. In general, they are functions
$\lambda_i$, with $i=1,2,..., 6$,which depend on the parameters and coordinates entering the tetrads $\vartheta^a$. 
As shown above, in the case of a vacuum solution the curvature matrix is traceless and hence the eigenvalues must satisfy the condition 
\be
\sum_{i=1}^6 \lambda_i =0 \ .
\label{id1}
\ee
In the case of a perfect fluid solution, the curvature eigenvalues are related by
\be
\sum_{i=1}^{6}\lambda_{i}= \frac{3\kappa}{2}(\rho + p) \ .
\label{id2}
\ee
These identities are a consequence of applying Einstein's equations to the general form of the curvature matrix ${\bf R_{AB}}$ and, consequently, they should contain information about the behavior of the gravitational field. We will verify this statements in the examples to be presented below.

\section{Repulsive gravity}
\label{sec:rep}

\begin{figure}[ht]
\includegraphics[scale=0.4]{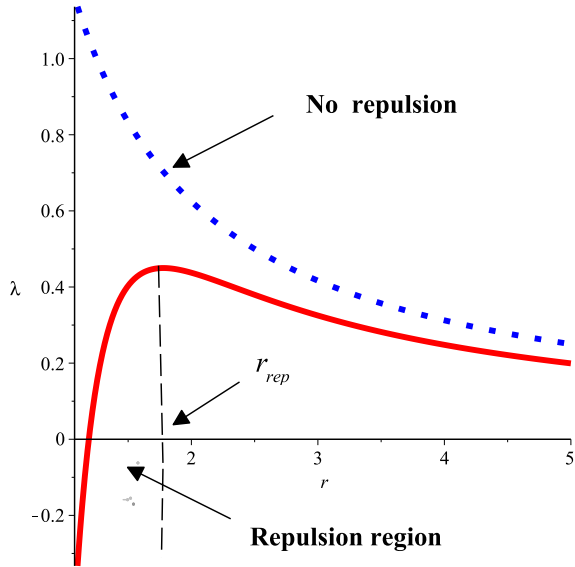}
\caption{Schematic representation of the behavior of the curvature eigenvalue. Here, $r$ represents the distance from the source.
\label{figrep}}
\end{figure}

Effects of repulsive gravity have been identified in the gravitational field of naked singularities and near black holes \cite{pqr13}.  
  In the literature, there are several intuitive definitions of repulsive gravity, but only recently an invariant definition was proposed in \cite{luongo2014characterizing} by using  the eigenvalues of 
the curvature tensor. The idea consists in using the eigenvalues to detect the regions of the gravitational field of compact objects, where repulsive effects are of importance. Indeed, since the gravitational field of compact objects is asymptotically flat, the eigenvalues should vanish at infinity. When approaching the object, the eigenvalues can either increase exponentially until they diverge at the singularity or 
they change their sign at some point, indicating the character of gravity has changed. This behavior is schematically illustrated in 
Fig. \ref{figrep}.
The dotted curve corresponds to an eigenvalue with no change in the character of gravity whereas the solid curve shows a maximum at a distance $r=r_{rep}$, indicating the presence of repulsive gravity. Within the region $r<r_{rep}$, repulsive gravity can be experienced and even becomes dominant once the eigenvalue changes its sign.

The above intuitive description of repulsive gravity can be formalized as follows. Let $\{\lambda_i^+\}\ (i=1,...,6)$ represents the set of eigenvalues of an exterior spacetime. As explained above, the presence of an extremum  in an eigenvalue is an indication of the existence of repulsive gravity. Then, let $\{r_l\},  l=1,2,... $ with $0<r_l<\infty$ represents the set of solutions of the equation
\be
\frac{\partial\lambda_i}{\partial r}\Big|_{r=r_l} = 0 \ , \quad {\rm with} \quad r_{rep} ={\rm max}\{r_l\} \ ,
\ee
i.e., $r_{rep}$ is the largest extremum of the eigenvalues and is called repulsion radius. Since the curvature eigenvalues characterize in an invariant way the gravitational interaction, the above definition represents a invariant method to derive the 
repulsion region of asymptotically flat spacetimes, which describe the gravitational field of compact objects.

To illustrate the applicability of the above definition, let us consider the Kerr-Newman spacetime \cite{stephani2009exact}: 
\begin{eqnarray}
ds^2 & = &  {\frac{r^2 - 2Mr + a^2+Q^2} {r^2 + a^2\cos^2 \theta}} (dt - a\sin^2\theta d\varphi)^2 
\nonumber \\ 
& & - {\frac{\sin^2 \theta} {r^2 + a^2\cos^2 \theta}}[(r^2 + a^2) d\varphi - adt]^2
\nonumber \\
 & &  - {\frac{r^2 + a^2\cos^2 \theta} {r^2 - 2Mr + a^2+Q^2}}dr^2 -
(r^2 + a^2\cos^2\theta)d\theta^2 \ ,
\label{kn}
\end{eqnarray}
where $M$ is the mass of the rotating central object, $a=J/M$ is the specific angular momentum, and $Q$ represents the total electric charge.
The orthonormal tetrad can be chosen as
\bea
\vartheta^0 = &   \left(\frac{r^2 - 2Mr + a^2+ Q^2} {r^2 + a^2\cos^2 \theta}\right)^{1/2} (dt - a\sin^2\theta d\varphi) \ ,   \nonumber\\
\vartheta^1 = &   {\frac{\sin\theta} {(r^2 + a^2\cos^2 \theta)^{1/2}}}[(r^2 + a^2) d\varphi - adt] \ , \nonumber\\
\vartheta^2 = &   (r^2 + a^2\cos^2 \theta)^{1/2} d\theta\ , \nonumber\\
\vartheta^3 = &    \left(\frac {r^2 + a^2\cos^2 \theta}{r^2 - 2Mr + a^2+ Q^2}\right)^{1/2} dr    \ .
\label{tetrad}
\eea
Following the approach presented in the last section, lengthy computations lead to the following eigenvalues 
\cite{luongo2014characterizing}
\be
\lambda_1^+ + i \lambda_4^+ = \lambda_2^++i\lambda_5^+  = l\ , \quad \lambda_3^+ +i \lambda_6^+ = -2l + k  \,,
\ee
where 
\be
l= - \left[ M-{\frac {{Q}^{2} \left( r+ia\cos  \theta
 \right) }{ {r}^{2}+{a}^{2}  \cos^2  \theta
^{2}}} \right]  \left( \frac { r -ia\cos  \theta   }{
  {r}^{2}+{a}^{2} \cos^2 \theta } \right) ^{3} \ ,
\label{funl}
\ee
\be
k = -{\frac {{Q}^{2}}{ \left( {r}^{2}+{a}^{2}  \cos^2  \theta
    \right) ^{2}}} \ .
\label{funk}
\ee
A straightforward analysis shows that all the eigenvalues have extrema located at different values of $r$, but the largest one is 
associated with the equation $\frac{\partial\lambda_3^+} {\partial r} =0$, the roots of which are determined by the equation 
\be
r^3(Mr - 2Q^2) + a^2\cos^2\theta [2Q^2 r + M(a^2 \cos^2\theta - 6 r^2)] = 0 \ .
\label{kncond}
\ee
Solutions of this equation can be obtained only numerically and depend on the values of $a$ and $Q$. However, analytical solutions can be obtained in the limiting cases. For  $a=0$, 
we obtain  the Reissner-Nordstr\"om repulsion radius
\be 
r_{rep}^{^{RN}} = \frac{2 Q^2}{M} \ ,
\ee
for $Q=0$, we get the Kerr repulsion radius of the Kerr source
\be
r_{rep}^{^K} = (1+\sqrt{2}) a \cos\theta\ ,
\ee
and in the Schwarzschild limiting case no repulsion radius is found. We see that the repulsion region is located very closed to the source. This agrees with the regions where repulsive effects have been identified by using a completely different approach based on the analysis of the motion of test particles along circular orbits around black holes and naked singularities \cite{pqr11a,pqr11b,pqr13}.

\section{$C^3$ matching}
\label{sec:mat}

The exterior field of compact objects is usually described by vacuum exact solutions characterized by singularities in a region closed to the source of gravity. To describe the entire spacetime and to avoid the presence of singularities, we usually say that the exterior spacetime must be matched with an interior spacetime which ``covers'' the region with singularities. To do this in concrete examples, we usually apply physical intuition to determine the matching surface and anyone of the $C^2$ methods mentioned in Sec. \ref{sec:int} to carry out the matching. The goal of the $C^3$ matching approach is to determine in an invariant manner where and how  to ``cover'' the singular spacetime region.

To be more specific, consider an exterior spacetime $(M^+, g_{\mu\nu}^+)$ and an interior spacetime $(M^-, g_{\mu\nu}^-)$, which are 
characterized by the curvature eigenvalues $\{\lambda_i^+\}$ and $\{\lambda_i^-\}$, respectively. Then, the $C^3$ matching procedure consists in $(i)$ establishing the matching surface $\Sigma$ as determined by the matching radius
\be
r_{match} \in [r_{rep}, \infty)  \ , \quad {\rm with} \quad r_{rep} ={\rm max}\{r_l\} \ , 
\quad \frac{\partial\lambda_i^+}{\partial r}\Big|_{r=r_l} = 0 \ , 
\ee
and $(ii)$ performing the matching of the spacetimes $(M^+, g_{\mu\nu}^+)$ and $(M^-, g_{\mu\nu}^-)$ at $\Sigma$ by imposing 
the conditions 
\be
\lambda_i^+\Big|_\Sigma = \lambda_i^-\Big|_\Sigma \quad \forall i\ .
\ee
Thus, the $C^3$ matching demands that the curvature eigenvalues be continuous across the matching surface $\Sigma$ which should be located
at any radius between the repulsion radius an infinity. Notice that the repulsion radius is determined by the eigenvalues of the exterior curvature tensor. Accordingly, the minimum matching radius coincides with the repulsion radius. Physically, this means that the $C^3$ matching is intended to avoid the presence of repulsive gravity in the case of gravitational compact objects.

\subsection{Newtonian gravity}
\label{sec:new}

Let us consider  the  line  element   for the nearly Newtonian metric in spherical coordinates
$x^{\alpha}=(t,r,\theta, \varphi)$	 (see \cite{misner2017gravitation}, page 470)
    \begin{eqnarray}\label{eq:Line_Element}
      ds^2 =   - (1 + 2 \Phi) dt^2 + (1 - 2 \Phi)(dr^2  +r^2 d\theta^2 + r^2 \sin^2\theta d\varphi^2) 
     \end{eqnarray}
     where  $\Phi<<1$ is the Newtonian potential. In this work, we limit ourselves to the study of spherically symmetric gravitational 
			configurations and so we assume that $\Phi$ depends  on $r$ only.   The components of the orthonormal tetrad are then 
\be
\vartheta^0 = \sqrt{1 + 2 \Phi}dt \ ,\quad
\vartheta^1 = \sqrt{1 - 2 \Phi} dr\ ,
\ee
\be
\vartheta^2 = \sqrt{1  {-} 2 \Phi}\, r \, d\theta \ ,\quad
\vartheta^3 = \sqrt{1  {-}  2 \Phi}\, r \sin\theta \, d\varphi\ ,
\ee
which in the first-order approximation lead to the connection 1-form  
\be
           \omega^0_{\ 1}   =   \Phi_r \vartheta^0,  \quad  
					\omega^2_{\ 3} = -\frac{1}{r} (1+\Phi) \cot{\theta} \vartheta^3,
\ee
\be
           \omega^1_{\ 2}   = -\frac{1}{r} (1+\Phi-r\Phi_r) \vartheta^2, \quad
            \omega^1_{\ 3}   = - \frac{1}{r} (1+\Phi-r\Phi_r) \vartheta^3 \ .
\ee
Moreover, the only non-vanishing components of the curvature 2-form can be expressed up to the first order in $\Phi$ as
\begin{eqnarray}
          \Omega^0_{\ 1}  & =  -  \Phi_{rr} \, \vartheta^0 \wedge \vartheta^1, \ \
          \Omega^0_{\ 2}   =     -  \frac{1}{r}   \Phi_r  \,  \vartheta^0 \wedge \vartheta^2, \\
           \Omega^0_{\ 3}  &  =   -  \frac{1}{r}   \Phi_r  \,  \vartheta^0 \wedge \vartheta^3, \ \
           \Omega^2_{\ 3}   =      \frac{2}{r}  \Phi_r \vartheta^2 \wedge \vartheta^3, \\
           \Omega^3_{\ 1}   & =      ( \Phi_{rr} + \frac{1}{r}  \Phi_{r} )   \vartheta^2 \wedge \vartheta^3, \ \
           \Omega^1_{\ 2}   =      ( \Phi_{rr} + \frac{1}{r}  \Phi_{r} )     \vartheta^1 \wedge \vartheta^2 \ .
\end{eqnarray}
It then follows that the only non-zero components  of  the curvature  tensor are 
\begin{eqnarray}
         R_{0 1 0 1}  &  = {\bf R}_{11} =     \Phi_{rr} \, , \ \
         R_{0 2 0 2}    = {\bf R}_{22} =   R_{0 3 0 3}    = {\bf R}_{33} =     \frac{1}{r}   \Phi_r  \, ,\\
         R_{2 3 2 3}  &  = {\bf R}_{44} =   \frac{2}{r}  \Phi_r  \, , \ \
         R_{3 1 3 1}    = {\bf R}_{55} =  R_{1 2 1 2}    = {\bf R}_{66} =    \Phi_{rr} + \frac{1}{r}  \Phi_{r}  \, .
         \end{eqnarray}
 Consequently, the curvature matrix ${\bf R_{AB}}$ (\ref{eq: CurvatureTensor}) is diagonal with eigenvalues
            \begin{eqnarray}\label{eq: eigenvs_of_R_inN_V2}
              \lambda_{1}  & = \Phi_{rr}  \ ,  \ \
              \lambda_{2}   =    \lambda_{3}   =    \frac{1}{r} \Phi_r\ , \\ 
              \lambda_{4} &  = \frac{2}{r}  \Phi_r\ , \ \
              \lambda_{5}    =     \lambda_{6}  =    \Phi_{rr}  +  \frac{1}{r} \Phi_r  \ ,
             \end{eqnarray}
which satisfy the relationship
\be
\sum_{i=1}^{6}\lambda_{i}= 3\left( \Phi_{rr} + \frac{2}{r} \Phi_r\right) =  3\ \nabla^2 \Phi \ .
\ee
We then conclude that in Newtonian gravity the eigenvalue identity (\ref{id2}) is equivalent to the Poisson equation, i.e.,
\be
\sum_{i=1}^{6}\lambda_{i}= \frac{3\kappa}{2} \rho \quad \Leftrightarrow \quad  \nabla^2 \Phi = \frac{\kappa}{2} \rho \ .
\label{poi}
\ee
This result is compatible with the above approach for the determination of the eigenvalues since we used the field equations to represent 
the curvature matrix as in Eqs.(\ref{pf1})-(\ref{pf2}). 

\subsubsection{$C^3$ matching in Newtonian gravity}

To illustrate the $C^3$ matching approach in Newtonian gravity, let us consider a spherically symmetric solution 
Then, the corresponding exterior field should correspond to that of a sphere which is a solution of the Laplace equation
  with  
	\be
    \rho_{ext}=0 \ ,\qquad  \Phi_{ext} = - \frac{M}{r},
		\label{ext1}
	\ee
where $M$ is  a constant. 
		It is then straightforward to calculate the curvature eigenvalues which turn out to be 
\be 
	\label{eq: NewtonianExterior_Solution_Eigenv}
  \lambda^+_{{1}}    = - \lambda^+_{{4}}= - \frac{2 M}{r^3},\ \ \
                           \lambda^+_{{2}}    =  \lambda^+_{{3}}   =  - \lambda^+_{{5}}    =  - \lambda^+_{{6}} =  
													\frac{M}{r^3}.
\ee
The condition 
\be
\frac{d \lambda^+_{i}}{dr}\bigg |_{r_{match}} = 0 
\label{cond1}
\ee
does not offer any positive finite root for the minimum matching radius, indicating that the matching can be performed at any radius 
$r_{match} \in (0,\infty)$.  
 We proceed now to the second step and demand 
 the equality of the  eigenvalues across the matching surface, i.e.,
\be
\lambda^-_{i} \bigg|_{r_{match}}= \lambda^+_{i} \bigg|_{r_{match}} \ .
\ee
Using the expressions (\ref{eq: eigenvs_of_R_inN_V2}) for the eigenvalues of the interior solution, we obtain that at the matching radius the following conditions must be satisfied
\be
\lambda^-_{1} = \frac{\kappa}{2}\rho- \frac{2}{r}\Phi_r = - \frac{2M}{r^3} \ , \quad
\lambda^-_{2} = \frac{1}{r}\Phi_r = \frac{M}{r^3} \ , \quad
\lambda^-_{3} = \frac{1}{r}\Phi_r = \frac{M}{r^3} \ , 
\ee

\be
\lambda^-_{4} = \frac{2}{r}\Phi_r =  \frac{2M}{r^3} \ , \ \
\lambda^-_{5} = \frac{\kappa}{2}\rho- \frac{1}{r}\Phi_r = - \frac{M}{r^3} \ , \ \
\lambda^-_{6} = \frac{\kappa}{2}\rho- \frac{1}{r}\Phi_r = - \frac{M}{r^3}  \ , 
\ee
where we have used Poisson's equation (\ref{poi}) to replace the second derivative of $\Phi$. The above equations represent a system of three independent algebraic equations from which we obtain that the only compatible solution is
\be
\rho = 0 \quad {\rm at} \quad r=r_{match} \ .
\label{cond2}
\ee
This condition is in agreement with our physical intuition as we expect that the density vanishes at the matching surface.   Notice that 
this result does not make use of any particular interior solution and so it is valid, in general, for any spherically symmetric field in Newtonian gravity. Nevertheless, for the sake of concreteness, we  now consider some solutions of the Poisson equation which determine interior Newtonian fields and can be expressed as  \cite{binney2011galactic}
\be
\label{eq:Homogeneus_Spheres_Sol}
\rho_{HS} = \rho_{_{0}} ,  \qquad \Phi_{HS}  = - \frac{k  \rho_{_{0}}}{4}  \left(a^2 - \frac{r^2}{3}\right) \ ,
\ee
  \begin{eqnarray}
\label{eq:Plummer_Model_Sol}
                           \rho_{P} = \frac{6 M b^2}{k (r^2 + b^2)^{5/2}}, \ \ 																			\Phi_{P} =- \frac{M}{(r^2 + b^2)^{1/2}}  ,
\end{eqnarray}
                           \begin{eqnarray}
                           \label{eq: Isochrone_Potential_Sol}
                           \rho_{IP} = \frac{2 M b \left[ 3b^2 + 3 b (b^2 + r^2)^{1/2} + 2r^2 \right] }{k (b^2 + r^2)^{3/2}  \left[ b + (b^2 + r^2)^{1/2} \right]^3  }, \quad
                            \Phi_{IP} = - \frac{M}{b + (b^2 + r^2)^{1/2}},
                            \end{eqnarray}      
where $\rho_{_0}$, $a$ and $b$ are constants.  
These solutions are known as the homogeneous sphere, Plummer model and isochrone potential, respectively.
As we can see, in general, non of these solutions satisfies the matching condition $\rho=0$, in general. This means that strictly speaking none of
them can be matched with the exterior solution of a sphere (\ref{ext1}). 
 However, to illustrate the validity of the matching procedure,  consider, for instance, the eigenvalues of the Plummer model
                           \begin{eqnarray}\label{eq:Plummer_Model_Eigenv}
                           \lambda^-_{{1}}  & =   \frac{M (b^2 - 2r^2)}{(b^2 + r^2)^{5/2}},\\
                           \lambda^-_{{2}}  & = \lambda^-_{{3}} =   \lambda^-_{{4}}/2 = \frac{M}{(b^2 + r^2)^{3/2}},\\
                             \lambda^-_{{5}}  & = \lambda^-_{{6}}  = \frac{M (2 b^2 - r^2)}{(b^2 + r^2)^{5/2}}
                           \nonumber 
                            \end{eqnarray}
A straightforward computation of the conditions $\lambda^+_{i} = \lambda^-_{i}$ shows that the only possible solution is $b=0$, which coincides with the matching condition $\rho_P=0$.  Moreover, from the expressions for the eigenvalues we see that  
\be
\lim_{r\to\infty} \lambda^-_{i} = \lambda^+_{Ei}\ ,
\ee
indicating that the matching can be performed only at infinity. An analysis the interior solutions for the homogeneous sphere and the isochrone potential leads to similar results. This corroborates the validity of the $C^3$ matching conditions in Newtonian gravity.

\section{Spherically symmetric relativistic fields}
\label{sec:sph}

For the investigation of relativistic fields, we consider the general spherically symmetric line element
    \begin{eqnarray}\label{eq:Line_Element2}
      ds^2 =   - e^{\nu} dt^2 + e^{\phi}dr^2  +r^2 (d\theta^2 + \sin^2\theta d\varphi^2) 
     \end{eqnarray}
where $\nu$ and $\phi$   depend  on $r$ only. It then follows that the corresponding orthonormal tetrad can be chosen as 
\be
\vartheta^0 = e^{\nu/2}dt \ ,\quad \vartheta^1 = e^{\phi/2} dr\ ,  \quad 
\vartheta^2 = r d\theta \ ,\quad \vartheta^3 = r \sin\theta d\varphi\ . 
\ee
It is straightforward to compute the  connection 1-form 
\begin{eqnarray}
\omega^1_{\ 2}  = - \frac{1}{r} e^{- \phi/2} \vartheta^2\ , \quad
\omega^1_{\ 3}  = - \frac{1}{r} e^{- \phi/2} \vartheta^3\ , \nonumber\\
\omega^2_{\ 3}   = - \frac{1}{r} \cot{\theta} \vartheta^3\ , \quad
\omega^1_{\ 0}   = - \frac{\nu_r}{2 r}  e^{- \phi/2} \vartheta^4\ ,
\end{eqnarray}
and from here the curvature 2-form and the components of the curvature tensor, which can be expressed as
\begin{eqnarray}
R_{0101} =  {\bf R}_{11} = - \frac{1}{4}  (\phi_r \nu_r  - \nu_r^2 - 2\nu_{rr}) e^{-\phi}\ , \quad
R_{0202} = {\bf R}_{22} =  \frac{1}{2r} \nu_r e^{-\phi}\ , \nonumber\\
R_{0303} = {\bf R}_{33}  =  \frac{1}{2r} \nu_r e^{-\phi}\ , \quad
R_{2323} = {\bf R}_{44} = \frac{1}{r^2} (1 - e^{-\phi})\ ,\\
R_{3131} = {\bf R}_{55}   = \frac{1}{2 r} \phi_r e^{-\phi}\ , \quad
R_{1212} = {\bf R}_{66}  = \frac{1}{2 r} \phi_r e^{-\phi}\ .  \nonumber     
\end{eqnarray}
We then obtain the following eigenvalues for the curvature  tensor of an interior perfect fluid solution
            \begin{equation}\label{eq: eigenvs_of_R_inGR}
              \lambda^-_{1}   = {\bf R}_{11}, \ 
						  \lambda^-_{2}   =  {\bf R}_{22},  \ 
              \lambda^-_{3}   =  {\bf R}_{33},  
				\end{equation}
				\begin{equation}
              \lambda^-_{4}   =  -  \lambda^-_{1}   +  \frac{k ( \rho + p) }{2} , \ 
              \lambda^-_{5}   =  -  \lambda^-_{2}   +   \frac{k ( \rho + p)}{2} , \ 
              \lambda^-_{6}   =  -  \lambda^-_{ 3}  +   \frac{k ( \rho + p) }{2}   .  
             \end{equation}
						
Moreover, Einstein's equations can be expressed as
\be
	\nu_{rr} +\frac{1}{2} \nu_r^2 -\frac{\nu_r}{2r}(2+r\phi_r) -\frac{\phi_r}{r}
-\frac{2}{r^2}(1-e^\phi)=0\  ,
\ee
\be
	\kappa \rho = \frac{1}{r^2} [ 1 + e^{-\phi} (r\phi_{r}-1)],\ \
	\kappa p = - \frac{1}{r^2} [1- e^{-\phi}(1+r\nu_{r}) ] .
	\ee

\subsubsection{$C^3$ matching in general relativity}

To proceed with the matching, we consider the Schwarzschild solution 
\begin{eqnarray}\label{eq: Schwarzschild_Sol}
                     \rho_{S} = p_{S} =0\ , \ \ 
										\phi_{S} = - \nu_{S} = -\ln{\left(1 - {2M}/{r}\right)} .
                      \end{eqnarray}
as the only available spherically symmetric vacuum solution. The eigenvalues are as follows: 
          \begin{eqnarray}\label{eq: Schwarzschild_Eigenvalues}
                       \lambda^+_{{1}} = - \lambda^+_{{4}} =  - {2 M}/{r^3}, \
                        \lambda^+_{{2}}   =  \lambda^+_{{3}} =   - \lambda^+_{{5}} =  - \lambda^+_{{6}} =   { M}/{r^3}.
                        \end{eqnarray}     
The $C^3$ matching condition $d \lambda^+_{i} /d r =0$ does not lead to any repulsion radius and so the matching can be carried out 
within the interval $r_{match}\in (0,\infty)$, resembling the situation in the case of Newtonian gravity. If, in addition, we demand that the exterior (\ref{eq: Schwarzschild_Sol}) and interior eigenvalues (\ref{eq: eigenvs_of_R_inGR}) coincide on the matching surface, we obtain that the only solution is
\be
\rho =0 , \ p=0\ .
\ee
Again, this result is very consistent and corroborates in an invariant way our physical expectation of vanishing pressure and density on the matching surface.

In the Appendix, we include a series of spherically symmetric interior solutions which are known as Tolman $I$-$VIII$. For all the solutions of this class we computed the interior eigenvalues, which are also included in the Appendix. It is straightforward to show that none of these solutions can be matched with the exterior Schwarzschild metric.  

\section{Conclusions}
\label{sec:con}

In this work, we propose a new method for matching two spacetimes in general relativity. We demand that the curvature eigenvalues of the 
interior and exterior solutions be continuous across the matching surface. To determine the matching surface, we assume that the exterior spacetime is asymptotically flat and consider the behavior of the corresponding eigenvalues as the source is approached. A monotonous growth of the eigenvalues is interpreted as corresponding to the presence of attractive gravitational interaction throughout the entire space. On the contrary, if an eigenvalue shows local extrema and even changes its sign as the source is approached, we interpret this behavior as due to the presence of repulsive gravity. The repulsion radius is defined by the location of the first extremum ($C^3$ condition), which appears as the source is approached from spatial infinity. In turn, the repulsion radius is defined as the minimum radius, where the matching can be carried out, i.e., the matching surface can be located anywhere between the repulsion radius and infinity. This means that the goal of fixing a minimum radius for the matching surface is to avoid the presence of  repulsive gravity because it has not been detected at least in the gravitational field of compact objects. 

We tested the $C^3$ matching procedure in the case of spherically symmetric perfect fluid spacetimes in Newtonian gravity and in general relativity. Remarkably, our method leads to completely general results, independently of any particular solution of the field equations. 
In the case of Newtonian gravity, we obtain the general result that the matter density of the gravitational source should vanish on the matching surface. In general relativity, the same result applies for the pressure. These conditions are very plausible from a physical point of view and, therefore, establish the validity of the $C^3$ matching. We analyzed several particular examples of well-known spherically symmetric perfect fluid solutions and found out that, in general, it is not possible to satisfy the $C^3$ matching conditions. This result indicates that to obtain physically meaningful interior solutions, it would be convenient to start from Ansatz, which satisfy the matching conditions from the very beginning.

In this work, we limited ourselves to the study of spherically symmetric solutions so that the matching surface is easily identified as a sphere. However, it is possible to apply the $C^3$ matching method  to the case of axially symmetric spacetimes, which are more realistic as models for describing the gravitational field of astrophysical compact objects. Preliminary results show that in the case of metrics with quadrupolar moment, the repulsion radius depend on the angular coordinate so that the matching surface deforms and differs from an ideal sphere. In this case, the $C^3$ matching implies a detailed numerical analysis of the curvature eigenvalues. This work is in progress and will be presented elsewhere.

As presented here, the $C^3$ matching method has been specially adapted for the study of asymptotically flat spacetimes, which can be used to describe the gravitational field of astrophysical compact objects. However, it is also possible to consider other physical situations in which, for instance, cosmological models or collapsing shells are to be matched. We expect to investigate this type of configurations in future works.

\section*{Acknowledgments}
A.C.G-P. is thankful to the Departamento de Gravitaci\'on y Teor\'ia  de Campos (ICN-UNAM) for its hospitality  during his research fellowship. The authors would like to thank Cesar S.  Lopez-Monsalvo and Francisco Nettel for useful comments and discussions.
This work was partially supported  by UNAM-DGAPA-PAPIIT, Grant No. 111617,  by the Ministry of Education and Science of RK, Grant No. 
BR05236322 and AP05133630,  and  by  COLCIENCIAS, Grant No. 110277657744  (VIE-UIS 8863).

\appendix

\section{Spherically symmetric interior solutions}
 
In this Appendix, we present the class of Tolman exact interior solutions \cite{tolman1939static} and calculate the corresponding curvature eigenvalues.


\subsection{Tolman I: The Einstein universe  }
          
             {\bf Solution:}
                     \begin{eqnarray}\label{eq: EinsteinUniverse_Sol1}
                      & \phi_{EU}  = -\ln{ \left( 1 - {r^2}/{ \kappa^2 } \right) }, \quad
                      \nu_{EU}  = 2 \ln{(c)}, \nonumber\\
                       &    \rho_{EU} = \frac{3}{k \kappa^2}, \quad
                            p_{EU} =  -   \frac{\rho_{EU}}{3},
                      \end{eqnarray}
                      where $c$, $k$ and $\kappa$  are  constants.
											
             {\bf Eigenvalues:}
                      \begin{eqnarray}\label{eq:  EinsteinUniverse_Sol}
                       \lambda_{EU_{1}} & =  \lambda_{EU_{2}}= \lambda_{EU_{3}}=0,\\
                        \lambda_{EU_{4}} & =   \lambda_{EU_{5}}= \lambda_{EU_{6}} = { 1}/{\kappa^2}. \nonumber
                        \end{eqnarray}


\subsection{Tolman II: The Schwarzschild-de Sitter solution }
           
					{\bf Solution:}
                     \begin{eqnarray}\label{eq: SdS_Sol1}
                      & \phi_{SdS}  = -\ln{ \left(1 -2M/r - r^2/\kappa^2   \right) }\ ,  \nonumber\\
                      & \nu_{SdS}  =  \ln{\left[c^2 \left(1 -2M/r - r^2/\kappa^2 \right) \right]}\ , \nonumber\\
                    &  \rho_{SdS}  =  - p_{SdS} =   \frac{3}{k \kappa^2},
                      \end{eqnarray}
                 where $c$, $k$ and $\kappa$  are  constants.
          
						{\bf Eigenvalues:}
                      \begin{eqnarray}\label{eq:  SdS_Sol}
                       \lambda_{SdS_{1}} & = -  \lambda_{SdS_{4}}=  - \frac{ 2 \kappa^2 M  +  r^3 }{ \kappa^2 r^3 },\\
                       \lambda_{SdS_{2}} & =  \lambda_{SdS_{3}}=  -\lambda_{SdS_{5}}  =  -\lambda_{SdS_{6}}
                                                          =\frac{  \kappa^2 M - r^3 }{ \kappa^2 r^3 }.
                        \end{eqnarray}

 


 \subsection{Tolman III: Schwarzschild interior  solution  } 
          {\bf Solution:}
                              \begin{eqnarray}\label{eq: Schwarzschild_Interior_Sol} 
                              \phi_{SI}  & = -  \ln{ \left[1 - {2M r^2}/{\kappa^3 }\right]}\ , \nonumber\\
                              \nu_{SI}   & = 2 \ln{ \left[ \frac{3( 1 - {2M}/{\kappa})^{1/2} }{2} - \frac{ \left( 1 - {2 M r^2}/{ \kappa^3 }  \right)^{1/2} }{2}\right] }\ , \\
                              \rho_{SI} &   =  \frac{6 M}{k \kappa^3}\ , \quad
                                  p_{SI}   = \frac{ 6 M \left[ (   1 - 2 M r^2/\kappa^3 )^{1/2}  - (1- 2M/ \kappa)^{1/2}  \right] }
                                                            { k \kappa^3 \left[ 3  (1- 2M/ \kappa)^{1/2} -  (   1 - 2 M r^2/\kappa^3 )^{1/2}  \right] }\ , \nonumber
                               \end{eqnarray}
                               where $k$, $\kappa$ and $M$ are constants.
                     
													{\bf Eigenvalues:}
                                    \begin{eqnarray}\label{eq: Schwarzschild_Interior_Eigenv}
                                    \lambda_{SI_{1}} & =  \lambda_{SI_{2}} =   \lambda_{SI_{3}} =
                                                                     - \frac{ 2M ( 2M r^2  - k^3)^{1/2} }{   \kappa^3 \left[      ( 2M r^2  - k^3)^{1/2}   
                                                                       - 3 \kappa ( 2M  - \kappa )^{1/2}           \right]  } , \\
                                    \lambda_{SI_{4}} & =  \lambda_{SI_{5}} =   \lambda_{SI_{6}} =  2M/ \kappa^3 .\nonumber
                                    \end{eqnarray}


 \subsection{Tolman IV  }
             
						{\bf Solution:}
                           \begin{eqnarray}\label{eq:TolmanIV_Sol}
                           \phi_{IV}   & = \ln{\left[ \frac{1 +2r^2/A^2}{ (1 -r^2/\kappa^2) (1 + r^2/A^2) }  \right]}\ , \ \
                           \nu_{IV}       = \ln{ [B^2 (1 + r^2/A^2)]}\ ,\\
                           &   & \nonumber\\
                           \rho_{IV}   & = \frac{  3A^2( A^2 + \kappa^2)  + (7A^2 + 2\kappa^2) r^2 +  6r^4}{ k  \kappa^2 (A^2 + 2 r^2)^2 }\ , \ \
                           p_{IV}       = \frac{\kappa^2 - A^2 - 3r^2}{k  \kappa^2 (A^2 + 2 r^2)}\ . \nonumber
                            \end{eqnarray} 
																
                           {\bf Eigenvalues:}
                           \begin{eqnarray}\label{eq:TolmanIV_Eigenv}
                              \lambda_{{IV}_{1}}   & = \frac{ (\kappa^2 - 2 r^2  ) A^2 - 2 r^4}{  \kappa^2 (A^2 + 2r^2)^2}, \\
                              \lambda_{{IV}_{2}}   & =  \lambda_{{IV}_{3}} = \frac{  \kappa^2 - r^2 }{  \kappa^2 (A^2 + 2r^2)}, \nonumber \\
                               \lambda_{{IV}_{4}}   & = \frac{ \kappa^2+  A^2 +  r^2}{  \kappa^2 (A^2 + 2r^2)}, \nonumber \\
                               \lambda_{{IV}_{5}}   & =  \lambda_{{IV}_{6}} = \frac{A^4  + (\kappa^2 + 2  r^2) A^2  + 2r^4}{  \kappa^2 (A^2 + 2r^2)^2}\ .\nonumber
                            \end{eqnarray}


 \subsection{Tolman V }

                          {\bf Solution:}                            
                           \begin{eqnarray}\label{eq:TolmanV_Sol}
                            \phi_{V}           & =\ln{\left[ \frac{ 1+ 2K - K^2}{ 1 - (1 + 2K - K^2)(r/\kappa)^N } \right]}\ , \quad
                            \nu_{V}                = \ln{(B^2r^{2K})}\ , \\
                            \rho_{V}            & = \frac{2K - K^2}{ k (1 + 2K - K^2) r^2} 
                                                         + \frac{3 + 5 K - 2K^2} { k (1 + K)\kappa^2} \left(\frac{r}{\kappa} \right)^{M}\ , \nonumber\\
                                     p_{V}        &   =   \frac{ K^2}{ k (1 + 2K - K^2) r^2} 
                                                         -   \frac{1 +  2K} { k \kappa^2} \left(\frac{r}{\kappa} \right)^{M}\ , \nonumber
                            \end{eqnarray}      
                            where $ N=2(1 + 2K -K^2)/(1 + K)$ and $M=2K(1-K)/(1+K)$.
                          
													{\bf Eigenvalues:}
                           \begin{eqnarray}\label{eq:TolmanV_Eigenv}
                               \lambda_{{V}_{1}}   & = - \frac{ K\left[  2 K ( K^2 - 2 K  - 1)(r/\kappa)^N  + K^2 - 1\right] }{  (K + 1)(K^2 - 2K - 1)r^2},\\
                                \lambda_{{V}_{2}}  & = \lambda_{{V}_{3}} = - \frac{K \left[  1 + (K^2 - 2 K - 1)(r/\kappa)^N    \right]}{(K^2 - 2 K - 1) r^2}, \nonumber\\
                                 \lambda_{{V}_{4}} & = \frac{(K^2 - 2K -1)(r/\kappa)^N + K^2 - 2K}{(K^2 - 2 K - 1)r^2},\nonumber\\
                                  \lambda_{{V}_{5}}& = \lambda_{{V}_{6}} = \frac{N}{2r^2} \left(\frac{r}{\kappa}\right)^N. \nonumber
                            \end{eqnarray}


 \subsection{Tolman VI  }

{\bf Solution:}                            
                           \begin{eqnarray}
                           \label{eq:TolmanVI_Sol}
                          &  \phi_{VI}          = \ln{\left( 2 - K^2 \right)} \  ,  \quad     \nu_{VI}  = 2 \ln{ \left(  A r^{1-K} - Br^{1+K}  \right) }\ , \nonumber\\
                          &  \rho_{VI}          =  \frac{1 - K^2}{ 8 \pi(2 - K^2) r^2}\ , \\
                           &   p_{VI}           =   \frac{ (1 - K)^2 A - (1 + K)^2 B r^{2K}  }{ 8 \pi(2 - K^2) (A - Br^{2K}) r^2} 
                                                         -   \frac{1 +  2K} { 8 \pi \kappa^2} \left(\frac{r}{\kappa} \right)^{M}\ , \nonumber
                            \end{eqnarray}      
                            where $A, B$ and $K$ are  arbitrary constants.
                            
{\bf Eigenvalues:}

                           \begin{eqnarray}
                           \label{eq:TolmanVI_Eigenv}
                               \lambda_{{VI}_{1}}   & = \frac{K \left[ A(K - 1) - B (K+1)r^{2K}  \right]  }{ (K^2 - 2) (A - B r^{2K}) r^2 }\ ,\\
                                \lambda_{{VI}_{2}}  & =  \lambda_{{VI}_{3}} = \frac{- A(K - 1) - B (K+1)r^{2K}   }{ (K^2 - 2) (A - B r^{2K}) r^2 }\ , \nonumber\\
                                 \lambda_{{VI}_{4}} & = \frac{K^2 - 1}{(K^2 -1)r^2}\ ,\nonumber\\
                                  \lambda_{{VI}_{5}}& = \lambda_{{VI}_{6}} = 0\ .\nonumber
                            \end{eqnarray}


 \subsection{Tolman VII  }

                       {\bf  Solution:}                            
                           \begin{eqnarray}
                           \label{eq:TolmanVII_Sol}
                            \phi_{VII}    & = -\ln{\left(1 -r^2/\kappa^2 +4r^4/A^4 \right)}\ ,
                            \nu_{VII}                 = 2 \ln{  \left[ B \sin{ \left( \ln{T^{1/2}}\right)}   \right] }\ , \\
                            \rho_{VII}    &    = \frac{3 A^2 - 20 \kappa^2 r^2}{ 8 \pi \kappa^2 A^2}\, \nonumber\\
                              p_{VII}       &  =  \frac{ B  \cot{ (\ln T^{1/2})}  
                                                                         - A^2(A^4 - 4\kappa^2 r^2) \Big( 1 -r^2/\kappa^2 + 4r^4/A^4   \Big)^{1/2}}
                                                                      {8\pi \kappa^2 A^6 \Big( 1 -r^2/\kappa^2 + 4r^4/A^4   \Big)^{1/2} }
                            \end{eqnarray}
                            where $ B \equiv  \left[4\kappa^2 A^4 - 4r^2 (A^4 - 4\kappa^2 r^2) \right] $ , 
                         $T$ is  given by    $$ c \,T \equiv  \Big( 1 -r^2/\kappa^2 + 4r^4/A^4   \Big)^{1/2} + 2r^2/A^2 - A^2/(4 \kappa^2) $$                            
                            and $c$, $\kappa$ and $A$ are  arbitrary constants.
                        
												{\bf Eigenvalues:}
                           \begin{eqnarray}
                           \label{eq:TolmanVII_Eigenv}
                               4 \lambda_{{VII}_{1}}  & = (1 -r^2/\kappa^2 + 4r^4/A^4) \nonumber\\
                                   & \times \left[  \frac{\cot{ (\ln T^{1/2})}}{T} ( F T_{r} 
                                                                         + 2 T_{rr} ) - \frac{T_{r}^2}{T^2}\left(  2  \cot{ (\ln T^{1/2})}  + 1\right)
                                                                          \right]\ ,\\
                                                                 F & \equiv     \frac{ -2r/\kappa^2 + 16 r^3/A^4 } {1 -r^2/\kappa^2 + 4r^4/A^4},\nonumber\\
                            \lambda_{{VII}_{2}}   & = \lambda_{{VII}_{3}} = - (1 -r^2/\kappa^2 + 4r^4/A^4) \frac{ T_r   \cot{  (\ln T^{1/2}) }  }{ 2r T}\ ,  \nonumber\\  
                            \lambda_{{VII}_{4}}   & = 1/\kappa^2 - 4r^2/A^4\  , \nonumber\\  
                             \lambda_{{VII}_{5}}   & =        \lambda_{{VII}_{6}} =  1/\kappa^2 - 8r^2/A^4\ . \nonumber
                            \end{eqnarray}    
                            Here $T_r$ indicates derivative with  respect to $r$.


 \subsection{Tolman VIII }
                         {\bf Solution:}                            
                         
                           \begin{eqnarray}
                           \label{eq:TolmanVIII_Sol}
                            \phi_{VIII}            & = - \ln{  \left[  \frac{2}{(a - b)(a + 2b -1)} - \left( \frac{2 M}{r}  \right)^{a+ 2b -1}  
                                                            - \left( \frac{r}{\kappa}  \right)^{a-b}    \right] } \ ,  \\
                            \nu_{VIII}            &     = \ln{ \left( B^2 r^{2b}\right)}  -  \phi_{VIII} \ , \nonumber\\
                       {8\pi r^2}   \rho_{VIII}   &  =   1 - \frac{2}{(a-b)(a + 2b -1)} - (a+ 2b - 2) \left( \frac{2 M}{r} \right)^{a+2b-1}  \nonumber\\
                                                           & +  (a - b + 1) \left( \frac{r}{\kappa} \right)^{a-b} \ , \nonumber
                                 \end{eqnarray} 
                                                                 
                       \begin{eqnarray*}
                         p_{VIII}   &   =  
                                                    \Big[( a - 2)(a  + 2b - 1)(a -b)  \left( \frac{2 M}{r}  \right)^{a+ 2b - 1} \nonumber\\
                                                             & - (a+b +1)(a + 2b -1 )(a - b) \left( \frac{r}{\kappa}  \right)^{a-b} \\
                                                             &  - a^2 +(1- b) a + 2b^2 + 3b + 2 \Big / \Big[   {8\pi (a -b) ( a + 2b -1) r^2}   \Big]\ ,                                  
                            \end{eqnarray*}      
                            where  $a$, $b$  and $M$  are  constants  and  $b \equiv (a^2 - a - 2)/(3- a)  $ 
                         
												{\bf Eigenvalues:}
                           \begin{eqnarray}\label{eq:TolmanVIII_Eigenv}
                                 \lambda_{{VIII}_{1}} /D  +   4b^2 + 4b & = 
                                                                       (a + 2b - 1)(a^2- b^2)(a- 1) \nonumber\\
                                                                       & \times \left[ \left( \frac{2 M}{r}  \right)^{a+ 2b - 1} + \left( \frac{r}{\kappa}  \right)^{a-b} \right] \ ,\\
                                 \lambda_{{VIII}_{2}}/D      + 4b   &=
                                                                       (a + 2b - 1)(a - b)   \nonumber\\ 
                                                                      &  \times \left[  ( 1- a) \left( \frac{2 M}{r}  \right)^{a+ 2b - 1} + (a+ b)\left( \frac{r}{\kappa}  \right)^{a-b} \right]
                                                                    , \nonumber\\ 
                                  \lambda_{{VIII}_{3}}    & =    \lambda_{{VIII}_{2}} \  ,\nonumber\\  
                                 \lambda_{{VIII}_{4}}/ 2D  +  2  & = 
                                                                          (a + 2b - 1)(a - b)  \nonumber\\ 
                                                                          & \times \left[   \left( \frac{2 M}{r}  \right)^{a+ 2b - 1} \left( \frac{r}{\kappa}  \right)^{a-b} +1  \right]
                                                                         , \nonumber\\ 
                                 \lambda_{{VIII}_{5}/D}  & =   \lambda_{{VIII}_{6}} = 
                                                                           (b - a)   \left( \frac{2 M}{r}  \right)^{a+ 2b - 1}   \nonumber\\ &
                                                                           + (a + 2b -1)\left( \frac{r}{\kappa}  \right)^{a-b} \ ,\nonumber                                                                       
                            \end{eqnarray}                 
                 where  $$ D \equiv \frac{1}{2(a-b)(a + 2b -1) r^2}  \ .$$


\providecommand{\newblock}{}


\end{document}